# Flexible and tensile microporous polymer fibers for wavelength-tunable random lasing

Van Duong Ta,[a,*] Dhruv Saxena,[b] Soraya Caixeiro[b,c] and Riccardo Sapienza[b]

Polymer micro-/nanofibers, due to their low-cost and mechanical flexibility, are attractive building blocks for developing lightweight and flexible optical circuits. They are also versatile photonic materials for making various optical resonators and lasers, such as microrings, networks and random lasers. In particular, for random lasing architectures, the demonstrations to-date have mainly relied on fiber bundles whose properties are hard to tune post-fabrication. Here, we demonstrate the successful implementation of an inverted photonic glass structure with monodisperse pores of 1.28 µm into polymer fibers with diameter ranging from 10 to 60 µm. By doping organic dye molecules into this structure, individual fibers can sustain random lasing under optical pulse excitation. The dependence of lasing characteristics, including lasing spectrum and lasing threshold on fiber diameter are investigated. It is found that the lasing emission red-shifts and the threshold decreases with increasing fiber diameter. Furthermore, owing to the mechanical flexibility, the lasing properties can be dynamically changed upon stretching, leading to a wavelength-tunability of 5.5 nm. Our work provides a novel architecture for random lasers which has the potential for lasing tunability and optical sensing.

## 1. Introduction

Random lasers have received a great deal of interest due to their rich physical properties[1] and a wide range of applications[2] including cost-effective light sources,[3] speckle-free laser imaging,[4] and biosensing.[5, 6] Random lasers, different to conventional lasers, rely on a highly scattering medium rather than a usual cavity to trap light.[7] As a result, one of the advantages of random lasers is that they are relatively simple and inexpensive to fabricate. Random lasers have been realized from laser dye in biological tissues,[8] colloidal suspensions,[9] photonic glasses,[10] semiconductor powders,[11] nanowires,[12] quantum dots,[13] bundles of nanofibers,[14] photonic networks.[15]

Among the above architectures, polymer micro-/nanofibers are interesting because they are cost-effective (can be easily mass-produced using electrospinning) and mechanically flexible, with the potential to make flexible photonic devices.[16] To date, polymer fibers based random lasers have mainly relied on randomly self-assembled three-dimensional fiber bundles,[17-21] where light diffuses inside the bundle, or individual fibers which lase similarly to conventional Fabry-Perot or whispering gallery mode (WGM) lasers.[22-25] In both cases, the lasing properties are hard to tune post-fabrication.

A random laser requires a highly scattering medium to confine the light and an optical gain for amplification, to reach stimulated emission. It has been demonstrated that a porous structure such as an inverted photonic glass, made of monodisperse spherical voids in a higher index matrix, can provide strong light scattering.[5] Often monodispersity is hard to achieve, and instead, polydisperse porous polymer fibers can be more easily fabricated; for example Bognitzki *et. al.* reported nanostructured fibers via rapid evaporation of the solvent,[26] and Zhang *et al.* obtained porous nanofibers by phase separation.[27] However, the study of these fibers has been so far restricted to passive devices, with no optical gain, though they have great potential for random lasing.

In this work, an inverted photonic glass structure with monodisperse pores is successfully fabricated in polymer fibers with diameters from 10 to 60 µm. Fibers are doped with organic dye molecules, thus allowing them to individually operate as random laser sources under optical pumping. Furthermore, the lasing wavelength of a typical fiber is tuned over a range of 5.5 nm by modifying the optical scattering strength via mechanically stretching the fiber.

## 2. Results and discussion

### 2.1. Dye-doped microporous polymer fibers

The fiber random lasers are fabricated in a cost-effective and biocompatible organic polyvinyl alcohol (PVA) polymer. PVA is widely considered as a safe material for use in medical and dietary supplement products.[28] Additionally, PVA is environmental friendly owing to its biodegradability and high water solubility. In the field of nanotechnology, PVA is considered as an appropriate material for producing micro-/nanofibers that can be employed for tissue scaffolding and drug delivery.[29] Very recently, PVA was demonstrated as an excellent material for a laser cavity.[30] A fluorescent dye, Rhodamine B (RhB), was selected for the gain material because it has a high quantum yield, up to 100%.

[a] *Department of Optical Devices, Le Quy Don Technical University, Hanoi 100000, Vietnam.*
[b] *The Blackett Laboratory, Department of Physics, Imperial College London, London, SW7 2AZ, UK.*
[c] *Present address: School of Physics and Astronomy, University of St Andrews, St Andrews KY16 9SS, UK*
[*] *E-mail of corresponding author: duong.ta@mta.edu.vn*

*†Electronic Supplementary Information (ESI) available: [the optical microscope image of various fibers, SEM image of a fiber's cross-section, emission spectra of four different fibers under various pump pulse fluences, simulation of scattering cross-section of a single ellipsoid air void in polymer]. See DOI: 10.1039/x0xx00000x*

Furthermore, the organic dye is water-soluble making it suitable for doping in the PVA matrix.

The fabrication method combines two processes: direct drawing[24] and selective chemical etching,[5] as shown in Fig. 1. Firstly, a droplet of an aqueous mixture which contains PVA, RhB and polystyrene (PS) microspheres is dropped on a glass substrate (Fig. 1a). The droplet is left for several minutes to dry in ambient conditions allowing water evaporation. Next, a metal tip is immersed into the viscous droplet and vertically retracted to form a fiber. The final diameter of a fiber is related to the drawing speed, a higher drawing speed results in a smaller diameter. The formation of the fiber can be seen in (Fig. S1a, ESI†). It is noted that the fiber was drawn within around 1 min. before the droplet is completely dried. Then, the fabricated fiber is placed on a home-made substrate that has a gap area (Fig. 1b). The lack of substrate adhesion in the gap area facilitates the formation of fibers with a circular cross-section (Fig. S1b, ESI†). After drawing from the solution, the fibers have a direct photonic glass structure, i.e. they are composed of spherical latex particles in a polymer matrix, as shown in Fig. 1c. Finally, the microporous fibers, composed of air voids in the polymer matrix, are achieved by removing the PS microspheres using selective chemical etching in dimethyl carbonate (Fig. 1d). The diameter of the polymer fiber does not change after the etching process because the PVA matrix does not dissolve in the etching solvent. The final etching step increases the refractive index contrast and consequently reduces the transport mean free path $l_t(\lambda)$. For similar but bulk geometries, $l_t(\lambda)$ has been measured to be around 4 µm for an inverse photonic glass at the wavelength of 600 nm.[5] By using the above technique, microporous fibers with diameter from about 10 to 60 µm are obtained (Fig. S2, ESI†) which is 3-15 optical thicknesses (assuming $l_t(\lambda) = 4$ µm).

The fabricated microporous fibers extend for hundreds of micrometers while keeping a uniform diameter and porosity. Fig. 2 shows the scanning electron microscope (SEM) image of a typical microporous fiber. The diameter of the fiber is relatively uniform (~23 µm) over a length of hundreds of micrometers (Fig. 2a) and gradually changes for longer lengths. The microporous fibres have pores in the form of round holes (Fig. 2b). It is worth noting that the PS microspheres are closely packed so microporous structures are distributed uniformly from the fiber surface to the fiber center (Fig. S3, ESI†). As a result, the microporous fiber efficiently scatters light, which can be exploited for random lasing. In addition, the fiber is highly flexible and bendable to form a closed loop, and stretchable due to the mechanical flexibility of the PVA host material. Under mechanical stretching, the fiber diameter decreases by around a factor of 2 and the air pores change their shape from round to ellipsoidal with an eccentricity of around 2 (Fig. 2c).

The scattering strength of a fiber depends on the density of the pores (around 50 %) and the scattering cross-section of the air voids. The scattering cross-section of the pores depends on the pore shape and changes when the spherical void becomes an ellipsoid. In the limit of very large stretching the pore becomes a very eccentric and collapsed ellipsoid, with minimal scattering. Therefore, by stretching the fibre, the scattering mean free path ($l_s$) can be tuned dynamically.

### 2.2. Random lasing from microporous polymer fibers

The dye-doped microporous fibers work as excellent random microlasers. Under optical pulse excitation, light emission from the dye molecules is multiple scattered between the PVA - air interface inside the microporous fiber as sketched in Fig. 3a. The light inside the fiber is trapped long enough for efficient amplification to occur and scatters out from the fiber in all directions.[31] Fig. 3b shows the optical and photoluminescence (PL) images of a 38 µm-diameter fiber captured using a CCD camera. The light emitted originates primarily from the pumped volume. The emission is not guided well along the porous fiber beyond a few transport mean free paths, which is expected as the scattering prevents ballistic light propagation. The emitted intensity increases linearly with the pump pulse energy until the threshold energy is reached, which is the onset of lasing action. The transition from fluorescence to lasing emission is evident in the emission spectrum recorded while increasing the pump pulse fluence (Fig. 3c). At the lowest pumping intensity of 38 µJ mm$^{-2}$, the emitted light has a low and broad-spectrum, characteristic of spontaneous emission. At 71 µJ mm$^{-2}$, the emission intensity increases sharply and the spectrum begins to narrow, indicating the development of stimulated emission. Lasing emission is dominant for pump fluences of 81 µJ mm$^{-2}$ and higher, and is characterized by narrow and bright spectral peaks. Figure 3d shows the PL peak intensity as a function of the pump fluence, exhibiting a sudden increase in the emission intensity at 75.5 µJ mm$^{-2}$ (equivalent to 380 nJ per pulse) identified as the lasing threshold. The threshold energy is comparable to other polymeric random lasers previously reported[5] and of the same order of other random lasers based on the photonic glass geometry.[32]

The spectral narrowing is another signature of random lasing. Fig. 4a shows that the spectral linewidth of PL emission decreases gradually with increasing pump energy. The evolution of the full-width-at-half-maximum (FWHM) of the emission fiber versus the pump fluences is plotted in Fig. 4b. FWHM of the initial fluorescent emission is about 50 nm at 21 µJ mm$^{-2}$, decreases sharply to 28 nm, nearly half of its original value, at 38 µJ mm$^{-2}$ and further to 13 nm at 61 µJ mm$^{-2}$. At the threshold of 75.5 µJ mm$^{-2}$, the spectral linewidth is only 5.5 nm. The linewidth reaches its minimum of 2.86 nm at 81 µJ mm$^{-2}$ and maintains this value for fluence up to 128 µJ mm$^{-2}$.

### 2.3. Size-dependence of lasing characteristics

The lasing threshold is an important parameter to describe the random lasing action. An experimental study of the size-dependence of the lasing threshold for microporous fibers (the same pore size and varying fiber diameter) was investigated. It is found that fibers with smaller diameters exhibit higher lasing thresholds and this is especially evident for fibers with diameters $D < 30$ µm as shown in Fig. 5a. The lasing threshold as a function of fiber's diameter is plotted in Fig. 5b (in log scale). For the 11 µm-diameter fiber, the threshold is about 200 µJ mm$^{-2}$, which reduces to 140 µJ mm$^{-2}$ for $D = 15$ µm, and further down to 81 µJ mm$^{-2}$ for $D = 28$ µm. When the size increases 2.5 times from 11 to 28 µm, the threshold decreases a similar amount of 2.4 times. The lasing threshold is much less affected by their size for the fibers with $D > 30$ µm. Indeed, when the fiber's diameter increases from 33 µm to 58 µm (a factor of 1.8) the threshold decreases minimally, from 73 to 68 µJ mm$^{-2}$ (a factor of 1.1). Overall, it is suggested that a $D^{-0.65}$ relationship is expected for fiber diameter and lasing threshold.

Besides the decrease of the lasing threshold, the lasing spectrum also varies with increasing fiber size. It is found that the lasing emission maximum red-shifts with increasing fiber diameter (Fig. S4 and S5, ESI†). When the diameter of fibers changes from 11 to 58 µm, a red-shift of 8.7 nm is recorded. A red-shift of the lasing spectrum has been observed previously, where the pump spot is increased,[32] and attributed to several factors including the mode confinement and reabsorption of the gain medium.[33] The effect is similar in our experiments, the pump spot is kept constant but the fibre diameter is increased, leading to the red-shift of lasing emission. This wavelength shift offers an opportunity for lasing emission control at the fabrication stage, but it is nevertheless static.

### 2.4. Wavelength-tunable lasing emission

Dynamically tunable lasers are very important for technological applications.[10] In conventional lasers, the lasing emission is determined by a cavity. As a result, their lasing wavelength can be tuned by modifying the laser cavity length.[34] In random lasers, the lasing emission is usually determined by the interplay of scattering and gain, and it is usually close to the maximum of the gain curve. The lasing spectrum can be tuned by modification of the gain spectrum, for example, by increasing the absorption in a frequency band,[35] or by changing the scattering spectrum in resonant scattering media.[10] Herein, we demonstrate the wavelength tunability by mechanical stretching as shown in Fig. 6. A dynamical change in the scattering strength affects the lasing properties, as it changes the light trapping in the gain medium. Under stretching, the air voids in the fiber gradually change their shape from sphere to ellipsoid and the diameter of the fiber decreases (Fig. 6a-c), leading to a change in scattering cross-section (Fig. S6-8, ESI†). The lasing spectrum is determined by the condition of maximal gain, which comes from the stimulated emission cross-section and density of the dye, and by the strength of light trapping in the gain medium, which is related to the transport mean free path $l_t(\lambda)$, the smaller $l_t(\lambda)$, the better the light trapping.[10] Subsequently, the stretching modifies the scattering strength of the fiber and is mainly responsible for the blue-shift of the lasing emission from 591.5 nm to 586 nm (Fig. 6d-e), i.e. with wavelength tunability of 5.5 nm. The change of the overall diameter of the fiber due to stretching, i.e. 39 µm which reduces to 24 µm after stretching, contributes to the lasing wavelength shift for about 44% of the total shift (Fig. S5, ESI†), which is 2.4 nm, while the total tunability is 5.5 nm, dominated by the change in scattering strength. While the tunability range is not as large compared to other devices such as a flexible thin-film random laser[36], it is in line with other polymer-based fiber lasers.[25] An interesting direction for further study is reversible wavelength tunability. Although this was not observed in our samples because the strain applied was too large resulting in inelastic deformation, it could be realized by limiting the strain or by using more elastic materials for synthesizing the fibers.

## 3. Conclusions

We have demonstrated microporous PVA fibers that can be fabricated by extrusion and selective etching. By doping organic dyes into these fibers, they can act as excellent random lasers under optical pumping, with the measured lowest threshold for the largest fibres as expected by a simple diffusive model. Owing to the mechanical flexibility of the host material, the random laser can be stretched and the emission wavelength tuned over a range of 5.5 nm. The wavelength shift mechanism is ascribed to the modification of the scattering cross-section of the microporous structure. The microporous fiber random lasers are promising for a wide range of commercial applications such as bendable, bright, and wavelength-tunable light sources, low-coherent illumination, optical sensing and imaging. Alternative fabrication techniques such as ink-jet printing, could in future be employed for the mass fabrication of these microporous fibers.[37, 38]

### 4. Experimental

### 4.1. Aqueous mixture for the fabrication of microporous fibers

The mixture used for the fabrication of microporous fibers was prepared by subsequently mixing 500 µL aqueous poly(vinyl alcohol) (PVA) solution, 4 wt%, with 50 µL aqueous Rhodamine B (RhB) solution, 1 wt% and 200 µL aqueous suspension, 10 wt%, polystyrene (PS) microspheres. The 4 wt% PVA solution was obtained by magnetic stirring 0.8 g PVA powder ($M_w$ = 89000-98000, from Sigma-Aldrich) in 20 mL deionized water at 80 °C for 3 hours. The 1 wt% RhB solution was achieved by directly dissolving 0.2 mg RhB (≥95% dye, from Sigma-Aldrich) in 20 mL deionized water. The monodisperse PS microparticles suspension (diameter of 1.28 µm and a standard deviation of 0.04 µm) was purchased from micro Particles GmbH

### 4.2. Optical measurement

Microporous fibers are generally stored in dimethyl carbonate to reduce the oxidation of the dye molecules. They were removed from the solution and left to completely dry prior to optical characterization. Individual porous fibers were investigated by using a micro-photoluminescence (µ-PL) setup. The pumping source was a pulsed microchip Nd:YAG laser (from Teem Photonics) at 532 nm wavelength, 400 ps pulse duration. The pumping laser beam was directed through an objective lens (magnification of 20× of a Nikon Eclipse Ti-U inverted microscope) and collimated to a beam spot of ~80 µm in diameter to excite the porous fibers. The polarized direction of the pumping laser was parallel to the fiber long axis. An acousto-optic modulator was used to precisely control the pump energy. Emission from the porous fibers was collected through the same objective and coupled to a spectrometer for spectral recording. For the mechanical stretching experiment, a 10× magnification objective lens was used for a larger field of view, resulting in beam spot size of ~165 µm. The fiber was mechanically stretched by suspending between the stationary microscope stage and a moveable stage and by applying tension. All optical measurements were carried at room temperature and at ambient conditions.

## Acknowledgements

This research is funded by Vietnam National Foundation for Science and Technology Development (NAFOSTED) under grant number 103.03-2017.318.

## Conflicts of interest

There are no conflicts to declare

ARTICLE

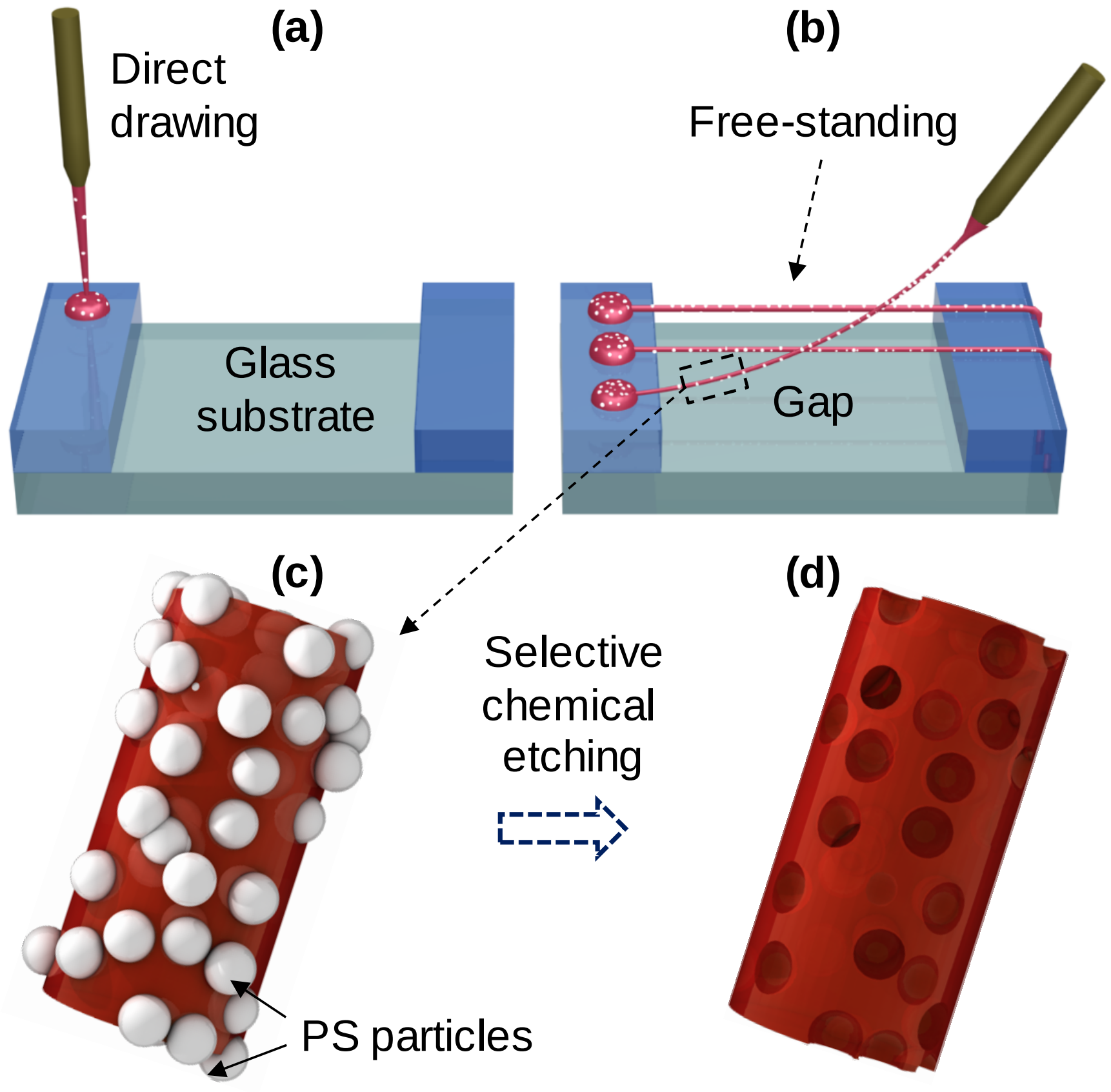

**Fig. 1.** Schematic diagram of the fabrication process of microporous fibers. (a) A fiber is directly drawn from an aqueous mixture. (b) The fabricated fibers are placed on a substrate with a gap area. Illustration of microfibers with (c) direct glass structure and (d) of microporous fibers (inverse structure) which are obtained by selective chemical etching in dimethyl carbonate.

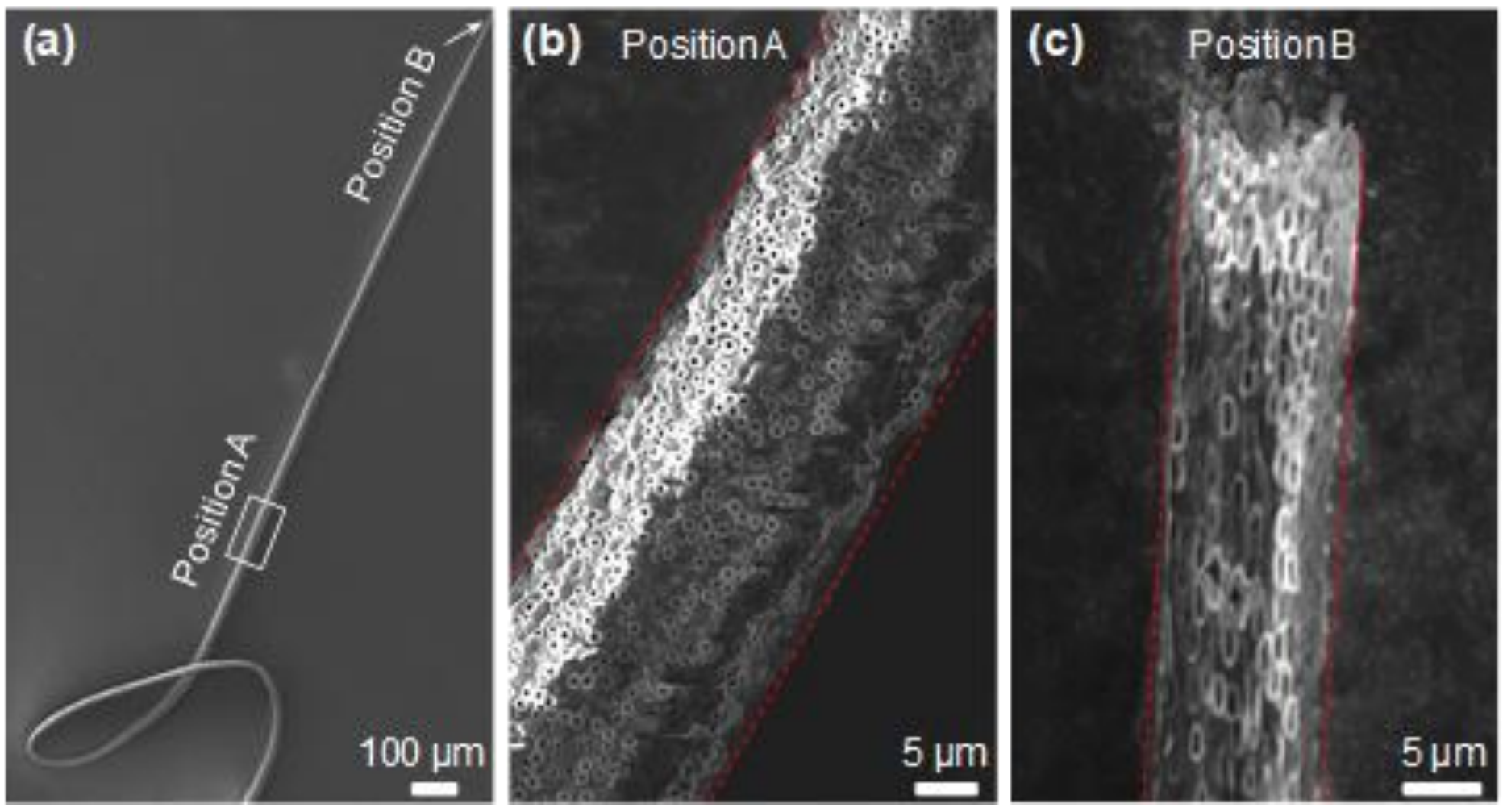

**Fig. 2.** (a) Scanning electron microscope (SEM) images of a typical porous microfiber. (b) and (c) High magnification SEM image of the fiber at the central and end part of the fiber, respectively. The fiber part in panel (b) is without stretching while the one in panel (c) is under mechanical stretching. The dashed red lines mark the edges of the fiber.

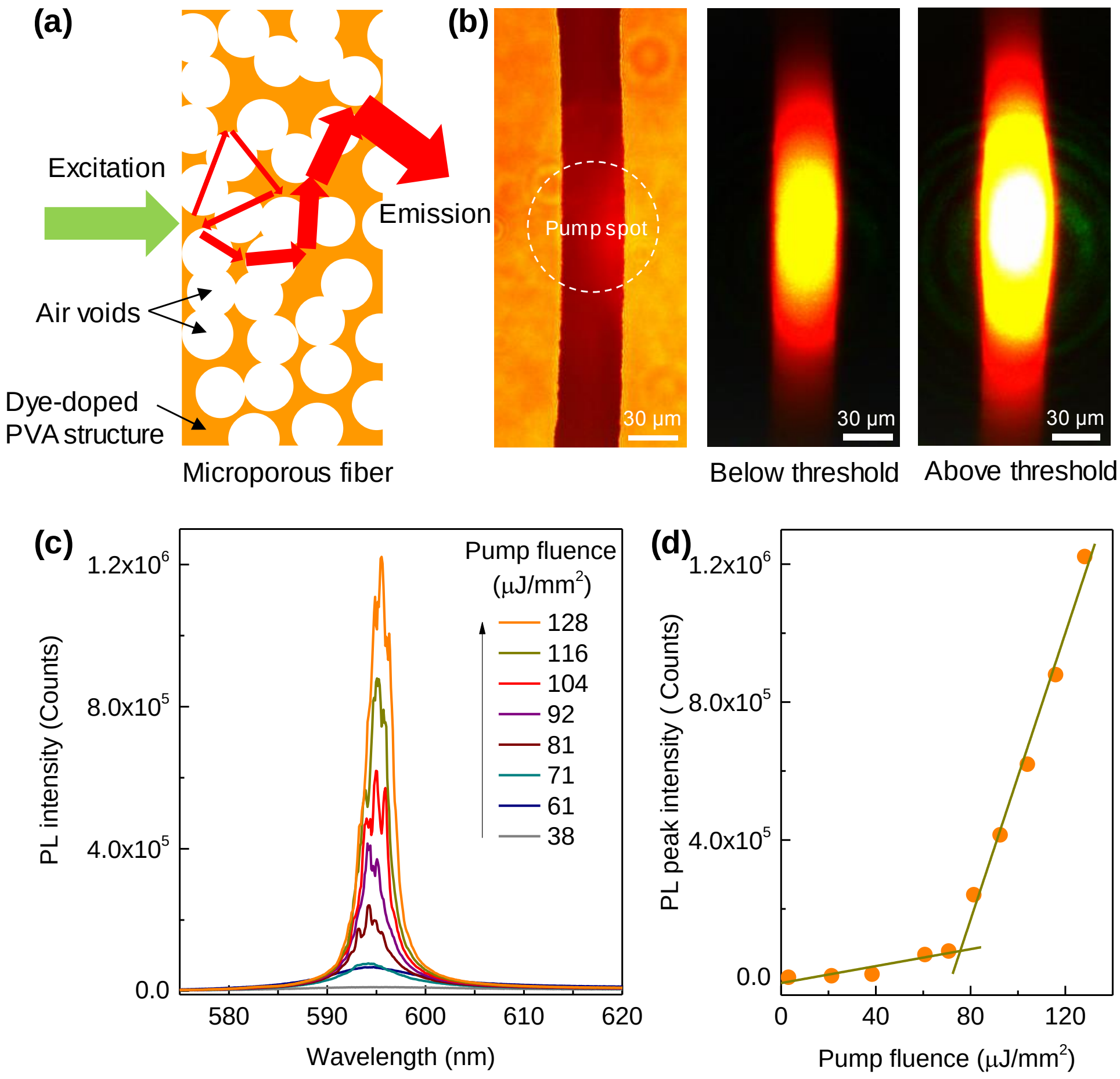

**Fig. 3.** (a) Schematic of light amplification via multiple scattering in a microporous fiber. (b) Optical microscope image of a typical porous fiber and its PL images under increasing pump energy. (c) Emission spectra of the fiber under various pump pulse fluences. (d) PL peak intensity of the fiber versus pump fluence exhibits a distinct lasing threshold.

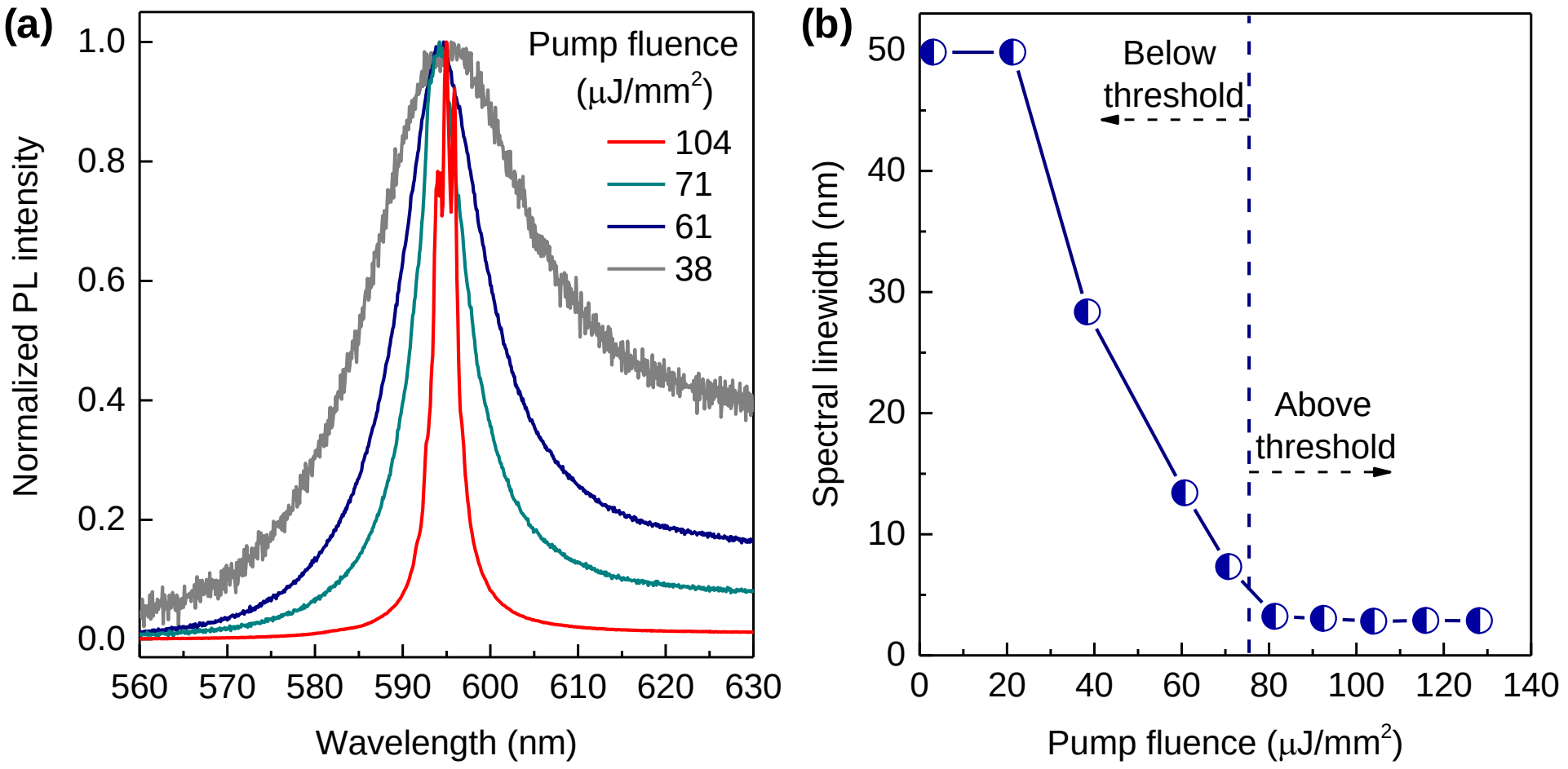

**Fig. 4.** (a) Normalized PL intensity of the fiber versus pump fluence. (b) Evolution of the FWHM of the emission versus the pump fluences. The vertical dash line displays the lasing threshold at about 75.5 μJ mm$^{-2}$.

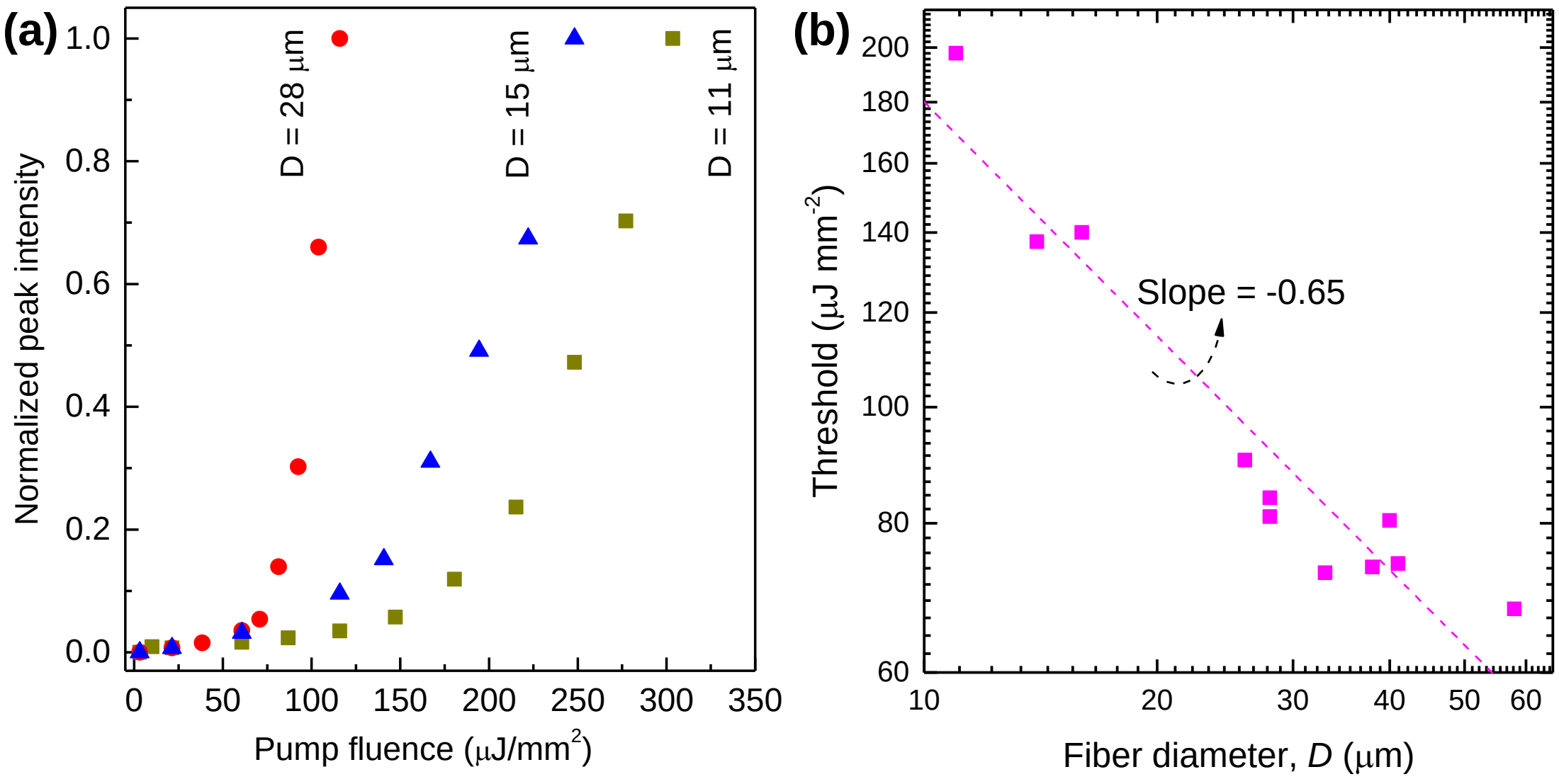

**Fig. 5.** (a) The relationship between peak intensity and pump fluence of three different fibers with diameter smaller than 30 μm. (b) Lasing threshold as a function of fiber's diameter. The dashed line is the linear fitted curve with a slope value of -0.65.

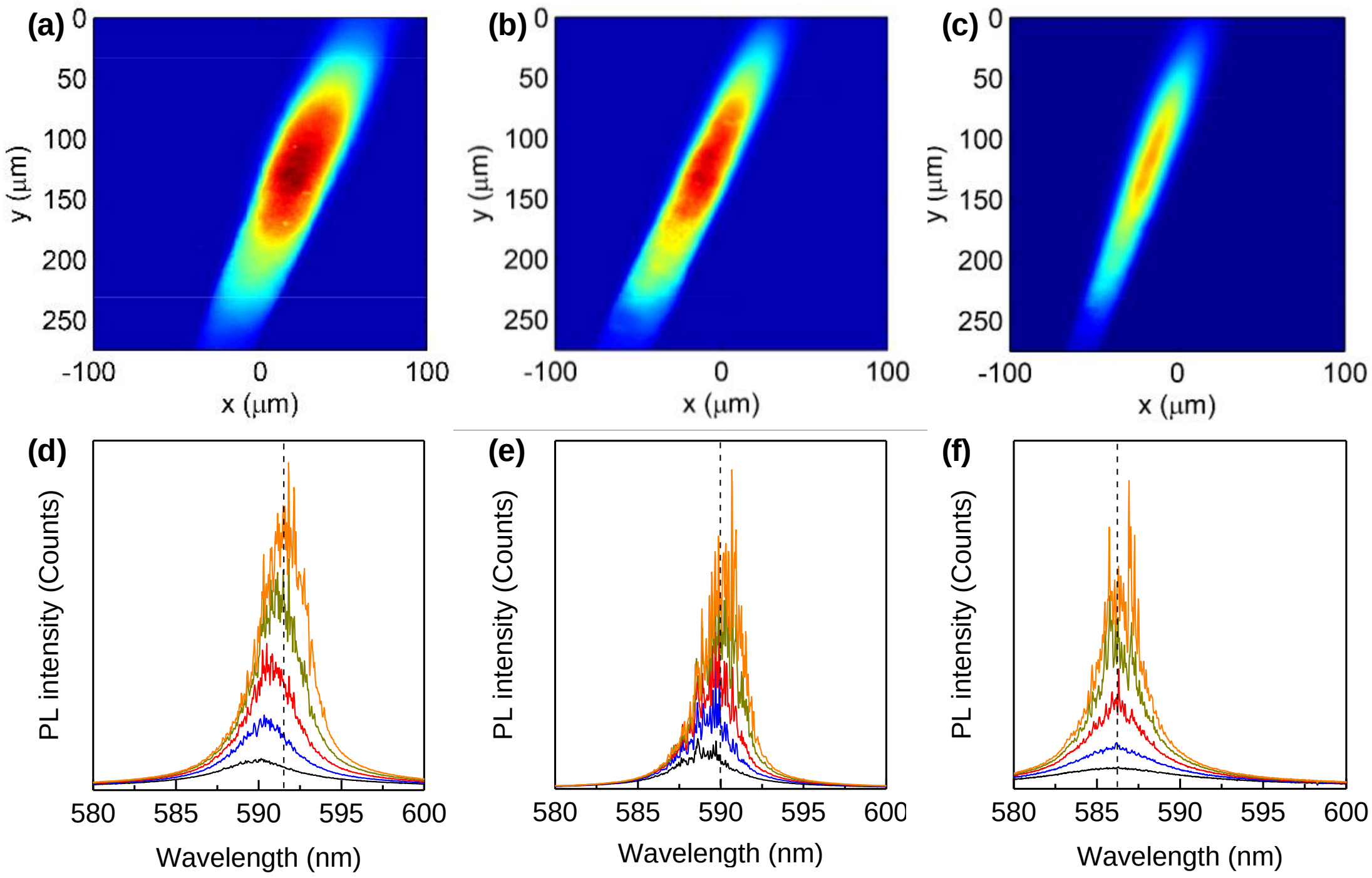

**Fig. 6.** (a)-(c) Images showing emission from an excited porous fiber (at fluence of 77 μJ mm$^{-2}$) under increasing mechanical stretching. (d)-(f) Emission spectra of the fiber with an increase of mechanical stretching and under various pumping fluences. The dash lines highlight the middle of the lasing spectrum

## TOC Figure

Flexible and tensile wavelength-tunable micrometer-sized random lasers in the form of microporous polymer fiber are demonstrated.

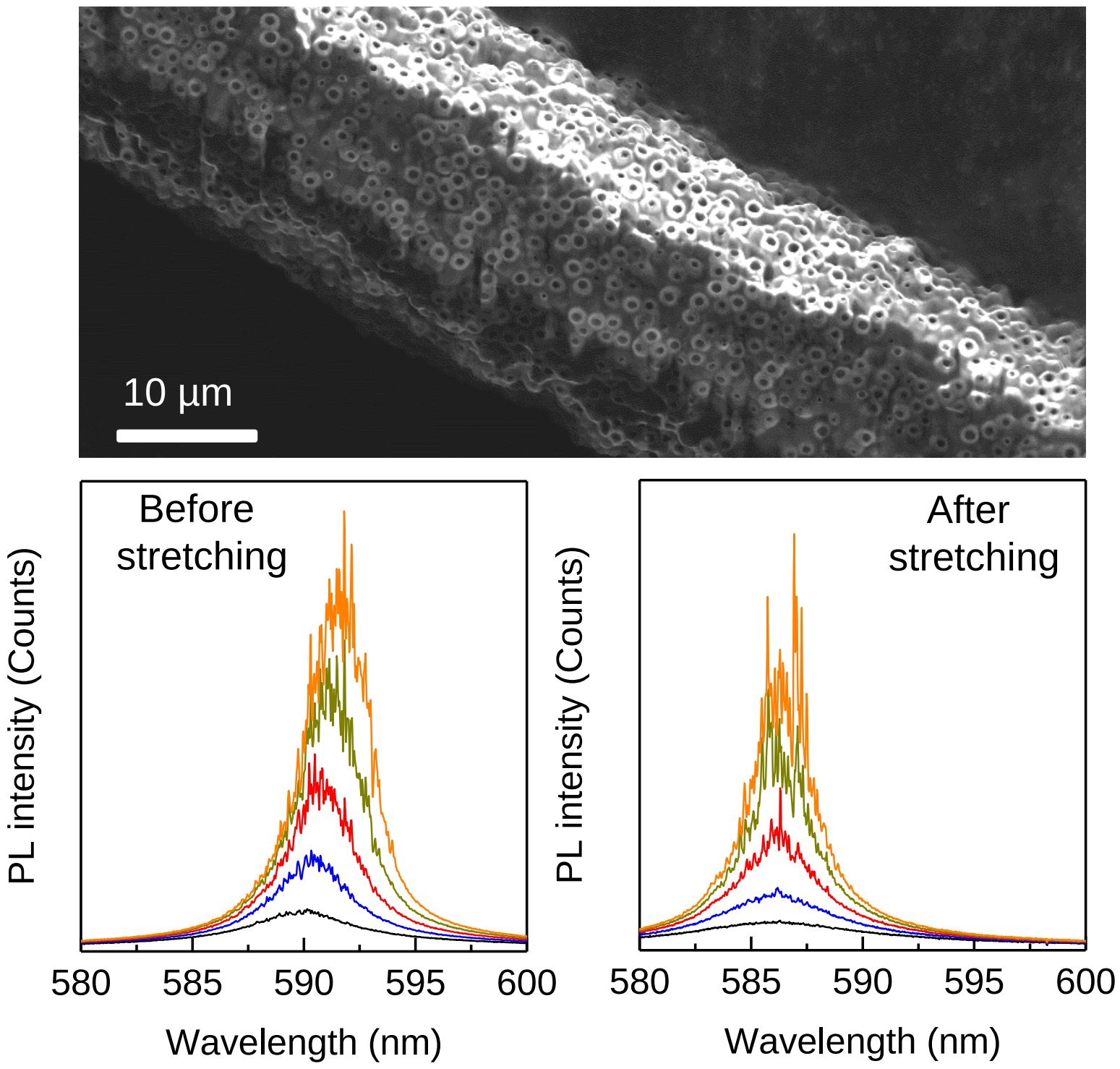

# Supplementary Information

## Flexible and tensile microporous polymer fibers for wavelength-tunable random lasing

Van Duong Ta,* Dhruv Saxena, Soraya Caixeiro and Riccardo Sapienza

PVA microfibers with different sizes can be fabricated by directly drawing from an aqueous mixture. Figure S1a shows the formation of a fiber created from an initial aqueous droplet. To study the fiber's cross-section, we embedded the fiber in Polydimethylsiloxane (PDMS, SYLGARD™ 184 Silicone Elastomer Kit, silicone base: curing agent = 10:1). When PDMS became solid, the fiber was cut vertically to its long axis and the cross-section was examined by using a microscope. The result shows that fabricated fibers have a nearly circular cross-section (Fig. S1b). Figure S2 shows that fibers with diameter from about 10 to 60 μm can be produced which is significant for the study of size-dependent lasing characteristics.

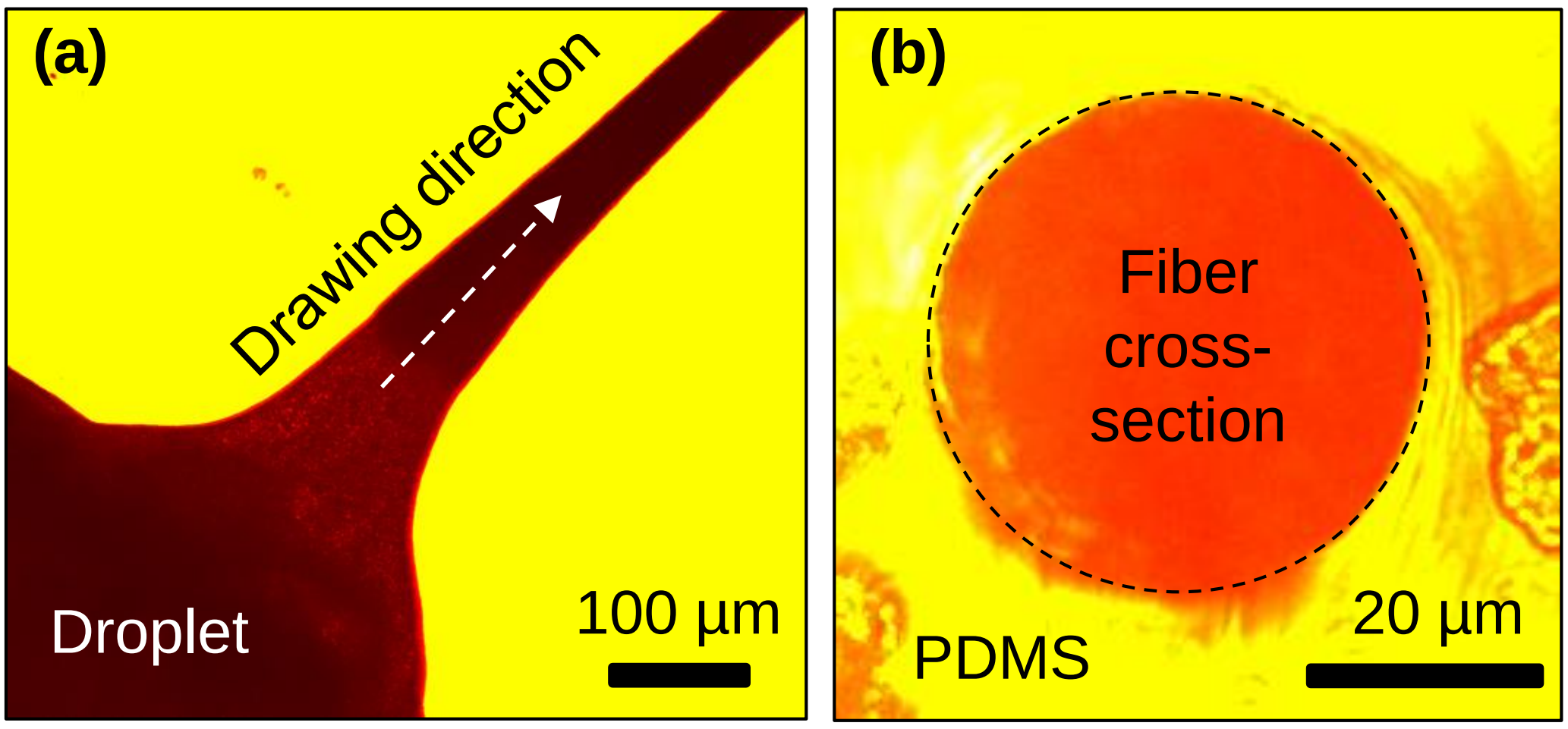

**Fig. S1.** Optical microscope image of (a) a direct drawing microporous fiber from a droplet and (b) a fiber's cross-section.

Microporous fibers were obtained by selective chemical etching in dimethyl carbonate. PS microspheres are completely removed, leaving round holes throughout the fibers. Figure S3 presents the SEM image of the cross-section of a typical fiber. It can be seen clearly that microporous are visible from the surface to the center.

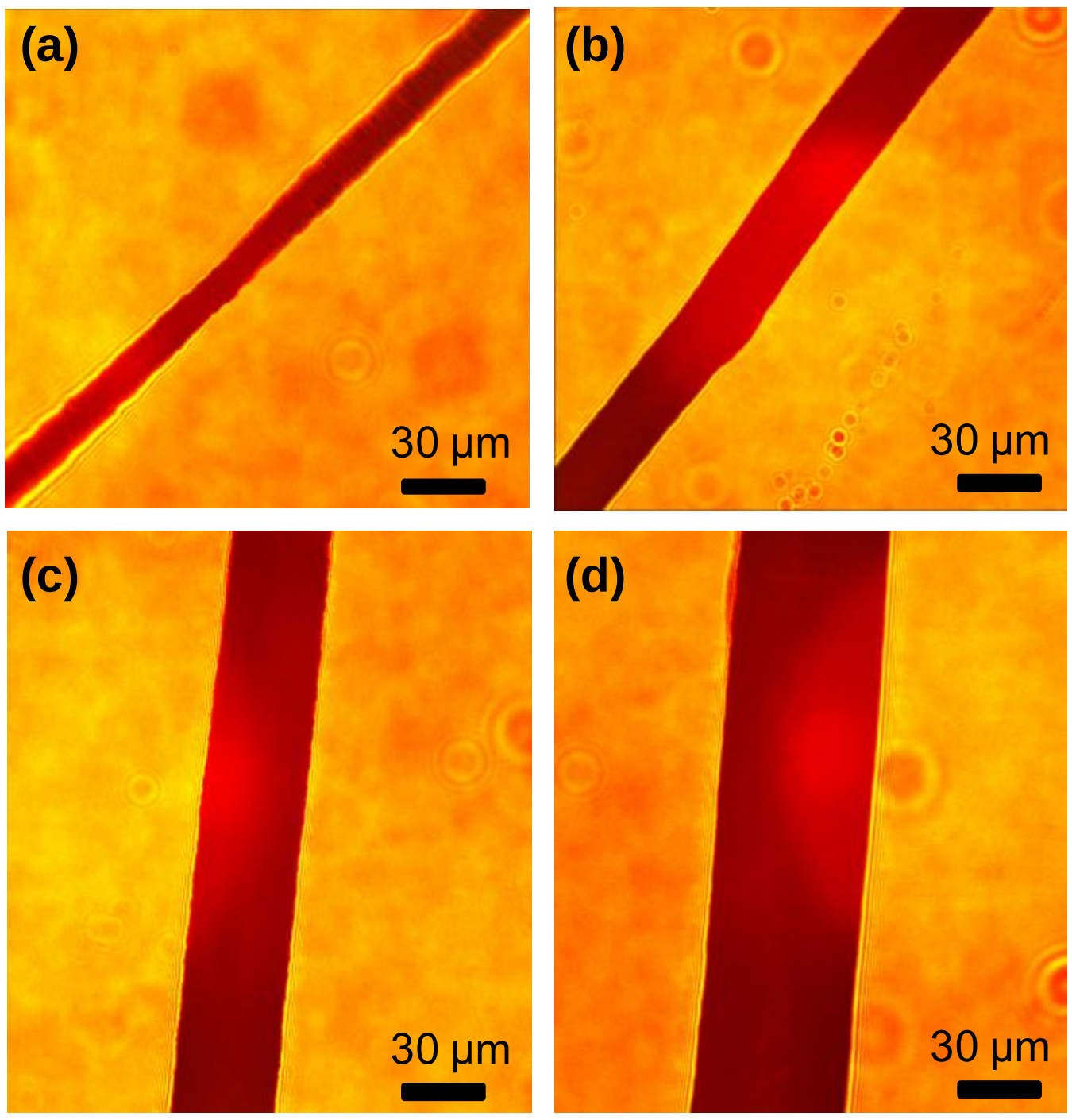

**Fig. S2.** Optical microscope images of different microporous PVA fibers. (a)-(d) Diameter of the fibers are around 14, 26, 38, 58 μm, respectively.

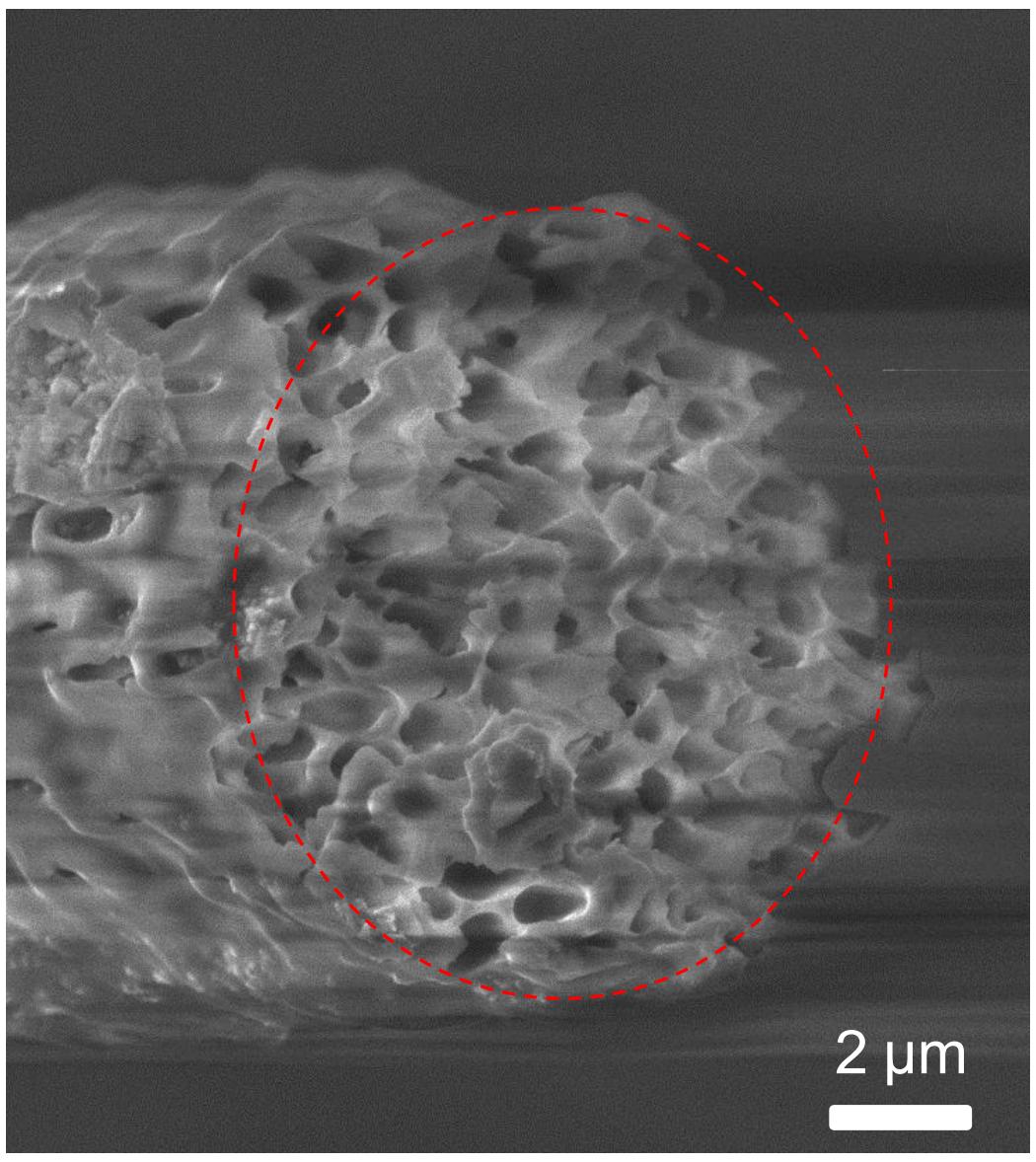

**Fig. S3.** SEM image of the cross-section of a microporous PVA fiber

Figure S4 shows size-dependent lasing spectra from microporous fibers. It can be seen that lasing spectra show a red-shift with the increase of fiber's diameter (*D*). The central wavelength of the 11 µm-diameter fiber is 587.7 nm (Fig. S4a). It gradually increases to 593 and 594.9 nm for 28 and 41 µm-diameter fibers, respectively (Figs. S4 b and S4c). When the diameter of fibers reaches 58 µm, a red-shift of 8.7 nm, from 587.7 nm to 596.4 nm, is recorded (Fig. S4d). Peak wavelengths of various fibers as a fuction of fiber diameter is shown in Fig. S5. The result suggests that if the diameter of a fiber is tuned then lasing wavelength can also be tuned in a specific range.

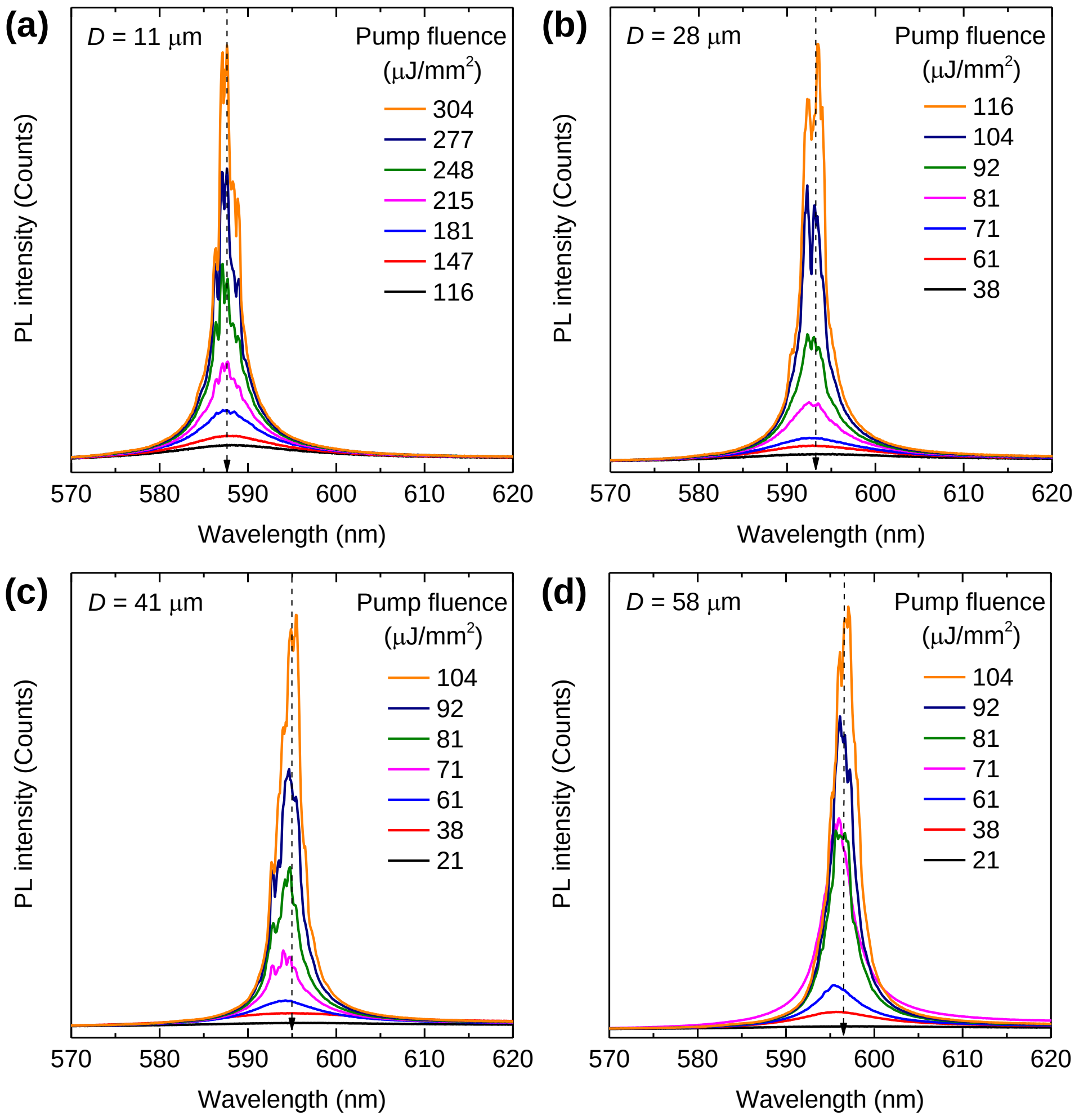

**Fig. S4.** Emission spectra of four different fibers under various pump pulse fluences. The dashed line denotes the center of the lasing spectrum.

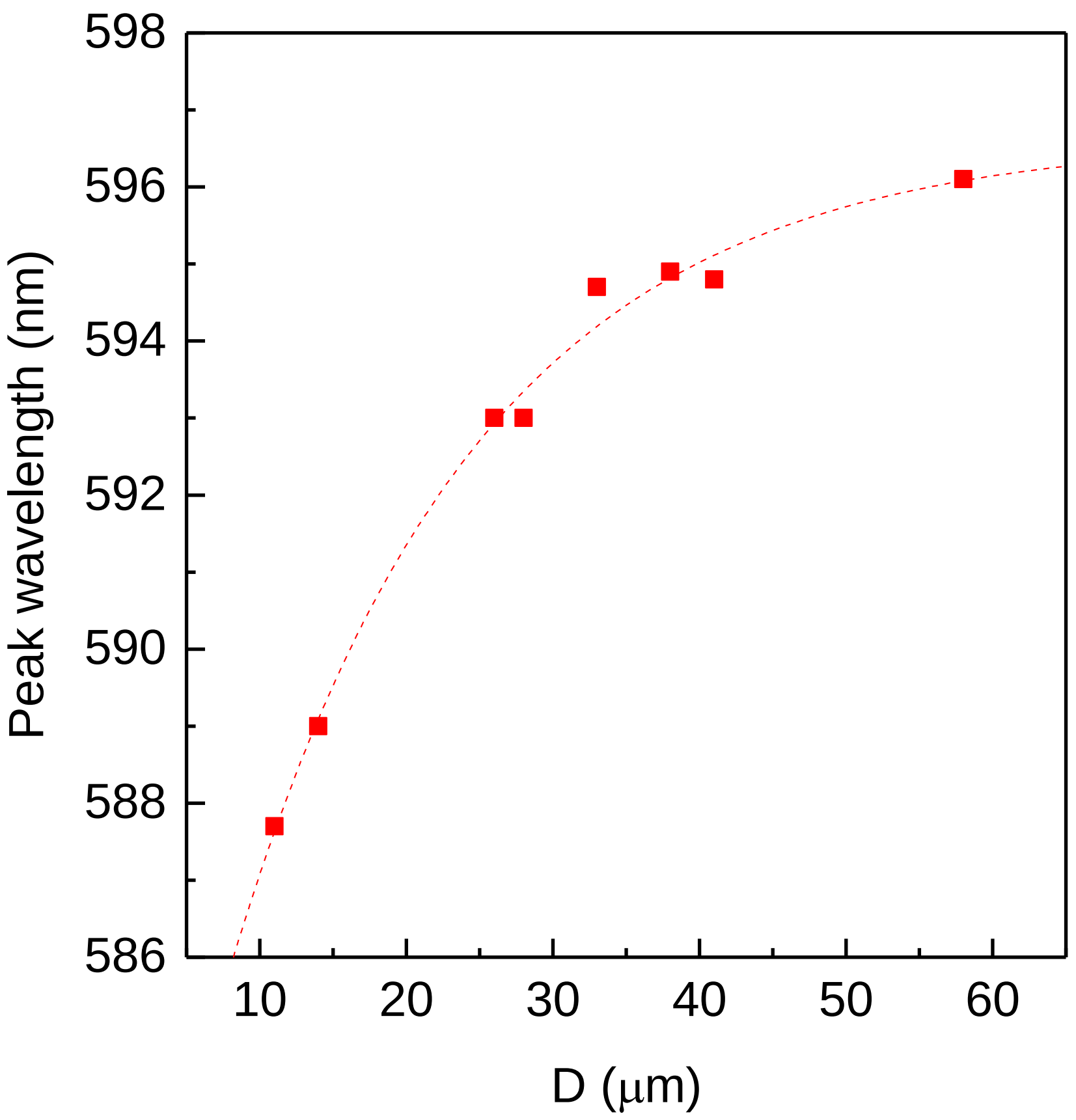

**Fig. S5.** Peak wavelength as a fuction of fiber diameter.

Figure S6 presents the simulation modes (using FDTD by Lumerical Inc.) for scattering cross-section of a single ellipsoid air void in polymer, considering all three different orientations with respect to the incident plane wave and the results are shown in Figure S7. It can be seen that the scattering resonance is red-shifted when the light is propagating parallel to the stretched direction and blue-shifted when it is transverse to the stretched pore. Overall, the average scattering cross-section changes with the stretching as shown in Figure S8 which would affect the output laser wavelength.

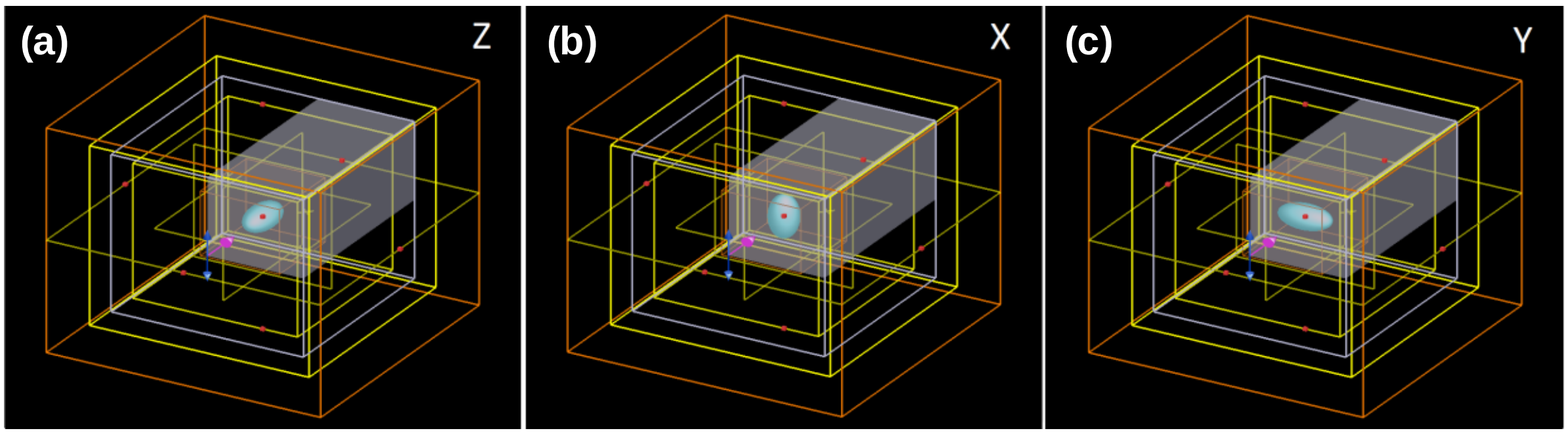

**Fig. S6.** Simulation geometry for calculating scattering cross-section of a single ellipsoid air void in polymer, considering all three different orientations with respect to the incident plane wave. The plane wave is propagating parallel to (a) the stretched direction and (b), (c) in transverse to the stretched pore. The incident light is polarized and the polarized direction is vertically shown by blue arrows.

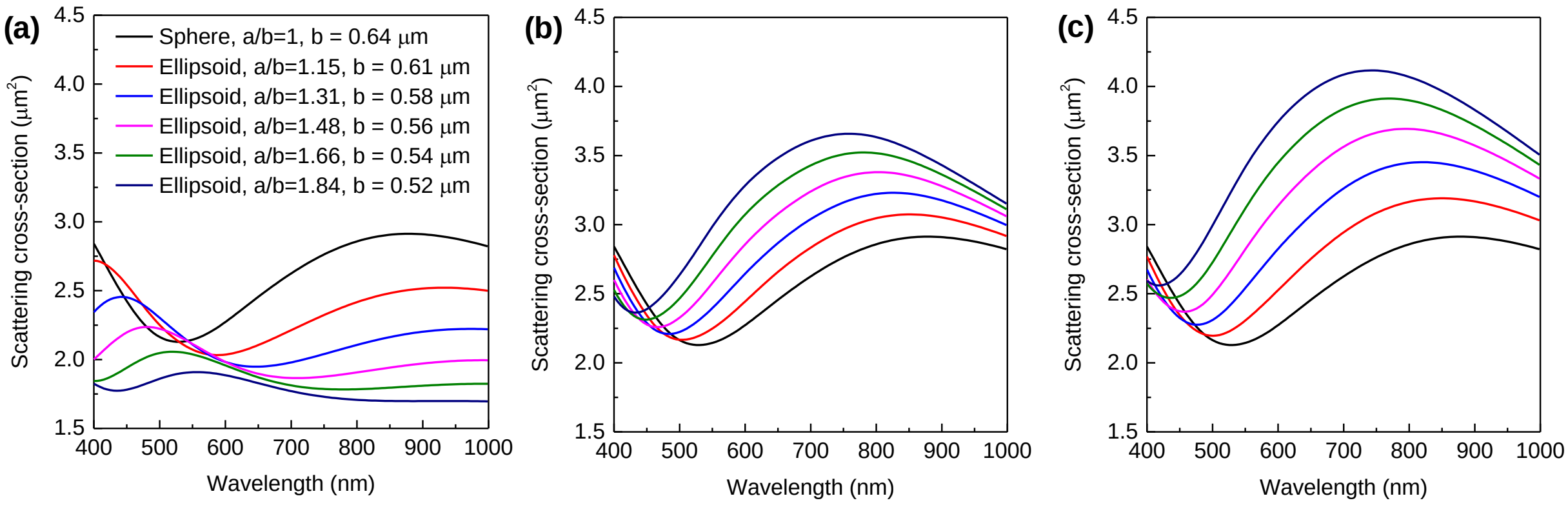

**Fig. S7.** Scattering cross-section of a single ellipsoid air void in polymer, considering all three different orientations concerning the incident plane wave. The plane wave is propagating: (a) parallel to the stretched direction and in (b), (c) transverse to the stretched pore.

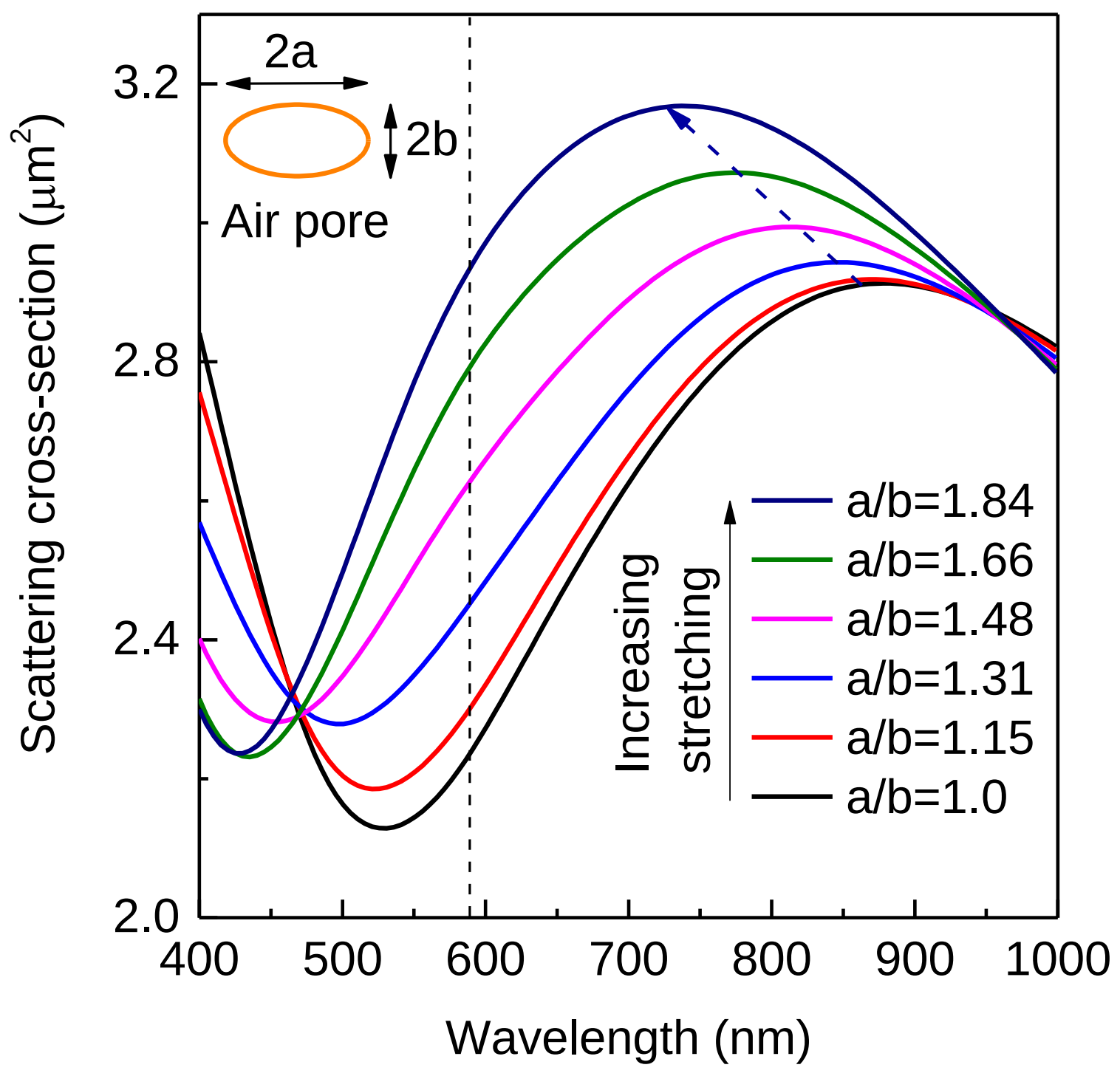

**Fig. S8.** Average scattering cross-section of air pore with various deformations in a polymer medium versus wavelength. The dashed line highlights the wavelength of 590 nm.